\documentclass[12pt,preprint]{aastex}

\shorttitle{Flares from Sagittarius A*}
\begin{document}


\title{Stochastic Electron Acceleration During the NIR and X-ray Flares in Sagittarius A*}


\author{Siming Liu,\altaffilmark{1} Fulvio Melia,\altaffilmark{2,3} and 
Vah\'e Petrosian,\altaffilmark{1,4}}

\altaffiltext{1}{Center for Space Science and Astrophysics, Department of Physics, Stanford 
University, Stanford, CA 94305; liusm@stanford.edu}
\altaffiltext{2}{Physics Department and Steward Observatory, The University of Arizona, 
Tucson, AZ 85721; melia@physics.arizona.edu}
\altaffiltext{3}{Sir Thomas Lyle Fellow and Miegunyah Fellow.}
\altaffiltext{4}{Department of Applied Physics, Stanford University, Stanford, CA 94305; 
vahe@astronomy.stanford.edu}


\begin{abstract}

Recent near-IR (NIR) and X-ray observations of Sagittarius A*'s spectrum have yielded
several strong constraints on the transient energization mechanism, justifying a
re-examination of the stochastic acceleration model proposed previously for these events. 
We here demonstrate that the new results are fully consistent with the acceleration of 
electrons via the transit-time damping process. But more importantly, these new NIR and 
X-ray flares now can constrain the source size, the gas density, the magnetic field, and 
the wave energy density in the turbulent plasma. Future simultaneous 
multi-wavelength observations with good spectral information will, in addition, allow us 
to study their temporal evolution, which will eventually lead to an accurate determination 
of the behavior of the plasma just minutes prior to its absorption by the black hole.

\end{abstract}


%% subject header.  Objects should be in the appropriate "individual"
%% headers (e.g. quasars: individual, stars: individual, etc.) with the
%% additional provision that the total number of headers, including each
%% individual object, not exceed six.  The \objectname{} macro, and its
%% alias \object{}, is used to mark each object.  The macro takes the object
%% name as its primary argument.  This name will appear in the paper
%% and serve as the link's anchor in the electronic edition if the name
%% is recognized by the data centers.  The macro also takes an optional
%% argument in parentheses in cases where the data center identification
%% differs from what is to be printed in the paper.

%\object{NGC 6624}, \objectname[M 15]{NGC 7078},
%\object[Cl 1938-341]{Terzan 8})}


\section{Introduction}

On 27 October 2000, the {\it Chandra} X-ray observatory detected a highly variable X-ray flare 
coincident with the position of Sagittarius A* (Baganoff et al. 2001). The transient lasted a couple 
of hours, with a peak luminosity $\sim 45$ times greater than the quiescent emission.  
Even more surprising 
was the realization that during the event, the X-ray output dropped abruptly by a factor of five in 
under ten minutes, recovering just as quickly.  Light travel-time arguments therefore place the 
source of this unusual radiation within a region no bigger than about 10 light-minutes or 
$\sim10^{13}$ cm across.

Sagittarius A*, a compact radio source at the Galactic center, is thought to be the 
radiative manifestation of a $\sim 3.4\times 10^6\;M_\odot$ supermassive black hole with 
the Schwarzschild radius $r_S\equiv 2GM/c^2\simeq10^{12}$ cm, where $G$, $M$ and $c$ are 
the gravitational constant, the black hole mass and the speed of light, respectively. 
Earlier theoretical modeling of its spectrum and polarization properties (Melia, Liu, 
and Coker 2000, 2001; see also Falcke and Markoff 2000 for an alternative model in which 
this emission is produced within a jet) had already anticipated an emission region 
within the inner ten Schwarzschild radii of a hot, magnetized Keplerian flow.

Not long after {\it Chandra's} first detection of the X-ray flare, XMM-{\it Newton} 
followed with its own measurements, including the discovery of two unusually 
strong bursts a couple of years later (Goldwurm et al. 2003; Porquet et al. 2003). None 
of the previous X-ray satellites had the sensitivity and spatial resolution to identify 
such low-luminosity events at the distance of the Galactic center. {\it Chandra} and 
XMM-{\it Newton} now detect them at a rate of about one per day, most of which are 
usually weak and last tens of minutes. The best-fit photon index during the majority of 
these bursts is $\sim 1.3$, representing a flattening of about 1 compared to Sagittarius 
A*'s spectrum in the quiescent state, which includes a significant contribution from 
thermal emission at large radii (Melia 1992; Baganoff et al. 2003). In addition, the 27 
October 2000 event appears to show a soft-hard-soft spectral evolution.

But the most intriguing X-ray flare of all may actually be the most recently detected 
event, in which an unambiguous modulation with an average period of $21.4$ minutes was 
seen over the course of its $\sim 3$ hour duration (Belanger et al. 2005). The 
separation between the flux minimums actually decreases from about $25$ down to $17.5$ 
minutes as the flare evolves, corresponding to the passage of an emitting plasma in a 
Keplerian motion from a radius $r\sim 2.9\;r_S$ to $2.4\;r_S$. This region 
therefore appears to lie somewhat below the marginally stable orbit (MSO) for a 
non-spinning black hole; it would, however, be outside the MSO for a Kerr black hole 
with a large spin. The monotonic decrease of the X-ray period is strongly reminiscent of 
what was seen in near-IR (NIR) flares detected just a few years earlier, where an average period of 
$17$ minutes was associated with a similar chirping behavior with the period decreasing from 23 to 
13 minutes (Genzel et al. 2003). These two sets of 
observations---one ground-based in the NIR, the other at X-ray energies from 
space---support the view that we are probably witnessing the evolution of an 
event moving inwards through the last portion of the accretion disk inside or very near 
the MSO. The inferred radial velocity $v_r\sim 10^8$ cm s$^{-1}$ is consistent with an 
accretion driven by the magnetic viscosity of the turbulent Keplerian flow.

Not surprisingly, previous speculation on the underlying mechanism for these events (Liu 
and Melia 2002) focused on the view that such X-ray flares might be driven by an 
accretion instability.  While this may still be true in light of all the more recent 
observations, the process by which the actual emission occurs is uncertain. However, the 
fact that the NIR spectrum (Eisenhauer et al.  2005) is much steeper than that of its 
X-ray counterpart excludes a direct extrapolation of the spectrum (see Liu and Melia 
2001 and Fig. \ref{fig2.ps} below). The currently favored scenario is one in which the 
mm/sub-mm to NIR portion of the spectrum is due to synchrotron, whereas the X-rays are 
produced by synchrotron-self-Compton (SSC).  (This constraint is empirically motivated, 
and is independent of whether the emission occurs within a disk or a jet.)
It turns out that producing the right blend of physical conditions to fit both the
NIR and X-ray flare emission (under the assumption that the two occur more or less
simultaneously---a concept that is yet to be confirmed compellingly) is not trivial.
In the next section we describe the observational constraints on the model parameters 
under this scenario. Our work with stochastic acceleration (SA) in producing 
Sagittarius A*'s quiescent spectrum (Liu, Petrosian, and Melia 2004, LPM04 hereafter; 
Liu, Melia, and Petrosian 2005) motivates us to consider a picture in which the flare 
itself is produced by a magnetic event, possibly driven by an accretion instability. A 
toy model of this acceleration and its fit to the flare spectra are described in \S\ 
\ref{model}. \S\ \ref{dis} summarizes the main results and discusses the model
limitations.

\section{Observational Constraints on the SSC Model}
\label{obs}

The NIR and X-ray flares in Sagittarius A* have peak luminosities as high as $\sim 
10^{35}$ ergs s$^{-1}$ (Baganoff et al. 2001; Ghez et al. 2004).  Several relatively 
long duration flares, two in the NIR (Genzel et al. 2003) and another in X-rays 
(Belanger et al. 2005), also displayed quasi-periodic modulations with a period 
decreasing from $\sim 25$ to $13$ minutes as the flare evolved.  Assuming Keplerian 
motion, this corresponds to a transition in radius from $\sim 2.9 r_S$ to $1.9 r_S$. Two 
flares have also been observed simultaneously in the NIR and X-ray bands, with a peak 
X-ray luminosity, respectively, 3 and 18 times the quiescent level of $\sim 2\times 
10^{33}$ ergs s$^{-1}$ and their spectroscopy still being processed (Eckart et al. 2004; 
Baganoff et al. 2005). The first NIR spectroscopy was completed in July 2004: a power-law fit to 
the power spectrum $\nu F_\nu\propto \nu^{-\alpha}$ (with $\nu$ the emission frequency) 
yields $\alpha=2.2\pm0.3$ during the peak of the flare observed July 15, 2004, and 
$\alpha=3.7\pm0.9$ during the rising and decay phases. For the flare of July 17, 2004, 
$\alpha=3.5\pm0.4$ (Eisenhauer et al. 2005). If the NIR emission is produced via the 
synchrotron process, the radiating electron distribution $N(\gamma)\propto \gamma^{-p}$ 
(with $\gamma$ the Lorentz factor) must have an index $p> 7.4$, suggesting an exponential cutoff of 
the electron distribution presumably dictated by the acceleration process, at 
$\gamma_{cr}\simeq(\nu_{\rm NIR}/\nu_B)^{1/2}$, where $\nu_B= eB_\bot/m_e c$ is the electron 
gyrofrequency, $e$, $m_e$ and $B_\bot$ are the electron charge, mass and the perpendicular magnetic 
field, respectively. The X-ray flares, on the other hand, often display a very hard spectrum with 
$\alpha\simeq -0.7\pm0.5$ (Baganoff et al. 2001; Goldwurm et al. 2003). In the SSC (or in general 
IC) scenario, this requires a flat electron spectrum ($p \simeq 1.6$). (This is different from a 
power-law commonly assumed, or a broken power-law distribution caused by radiative cooling.) At the 
longer sub-mm wavelength the flares usually have a much smaller flux increase above their quiescent 
value than the NIR flares, which suggests a photon spectral index $\alpha<2.2$, requiring a 
flattening of the electron distribution at lower energies. 
%These electrons may Compton-scatter the sub-mm to NIR photons to produce the hard X-ray spectrum. 
As we shall show below a fairly flat power-law spectrum with an exponential cutoff $N(\gamma)=N_0 
\gamma^{-p}\exp(-\gamma/\gamma_{cr})$ can reproduce these observed spectra and is a natural 
consequence of a simple SA model. Because most of the observed NIR 
and X-ray emissions are produced by electrons near $\gamma_{cr}$, and for $p<0.4$ the radiation 
spectrum is almost identical, we set $p=-2$, corresponding to a relativistic Maxwellian 
distribution, in what follows. The observed emission characteristics can then set strict limits on 
the model parameters, such as the source size $R$, the magnetic field $B$, the gas density $n$, and 
$\gamma_{cr}$.

\begin{figure}[bht] 
\begin{center}
\includegraphics[height=8.4cm, width=8.0cm]{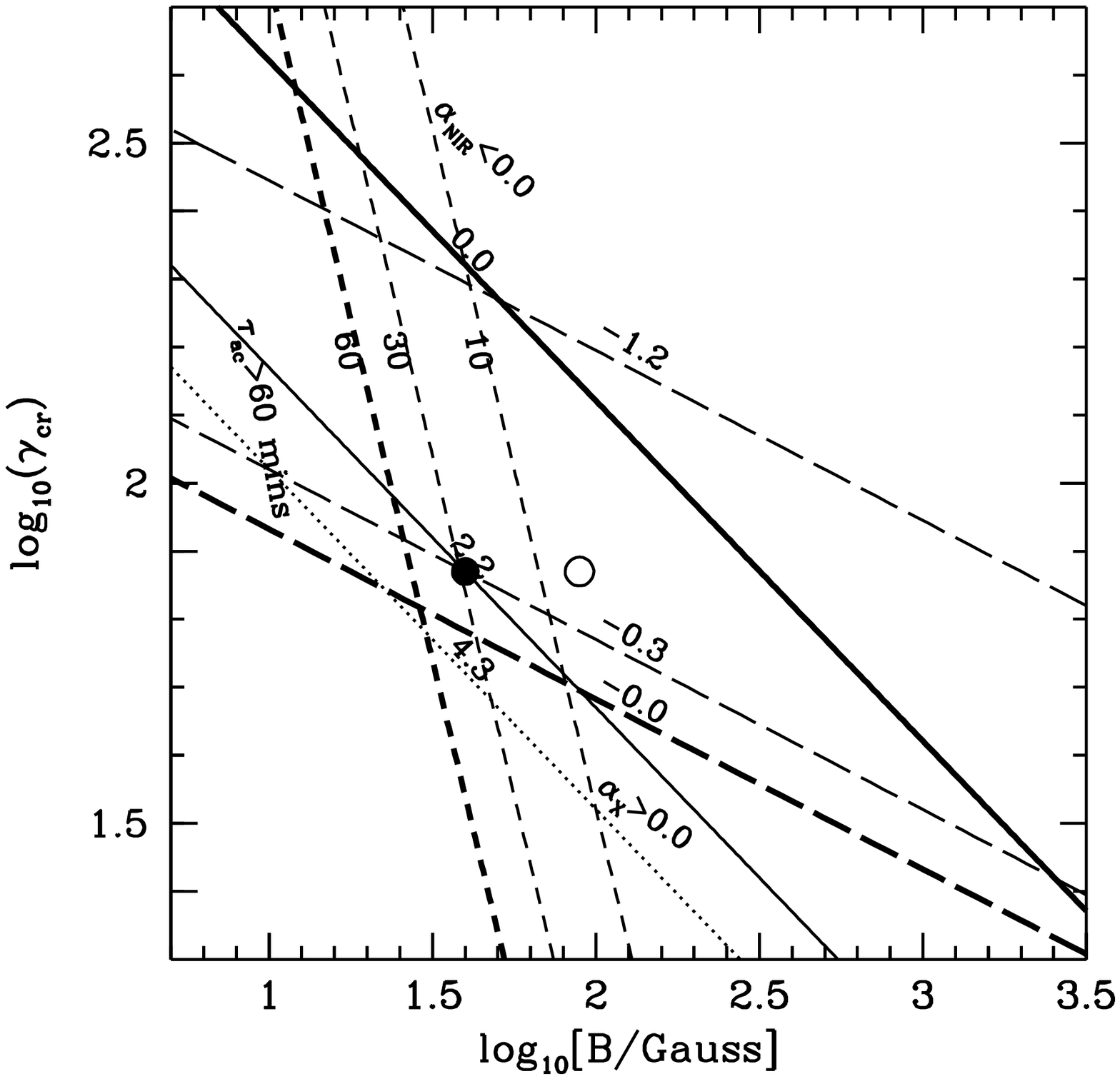}
\hspace{-0.6cm}
\includegraphics[height=8.4cm]{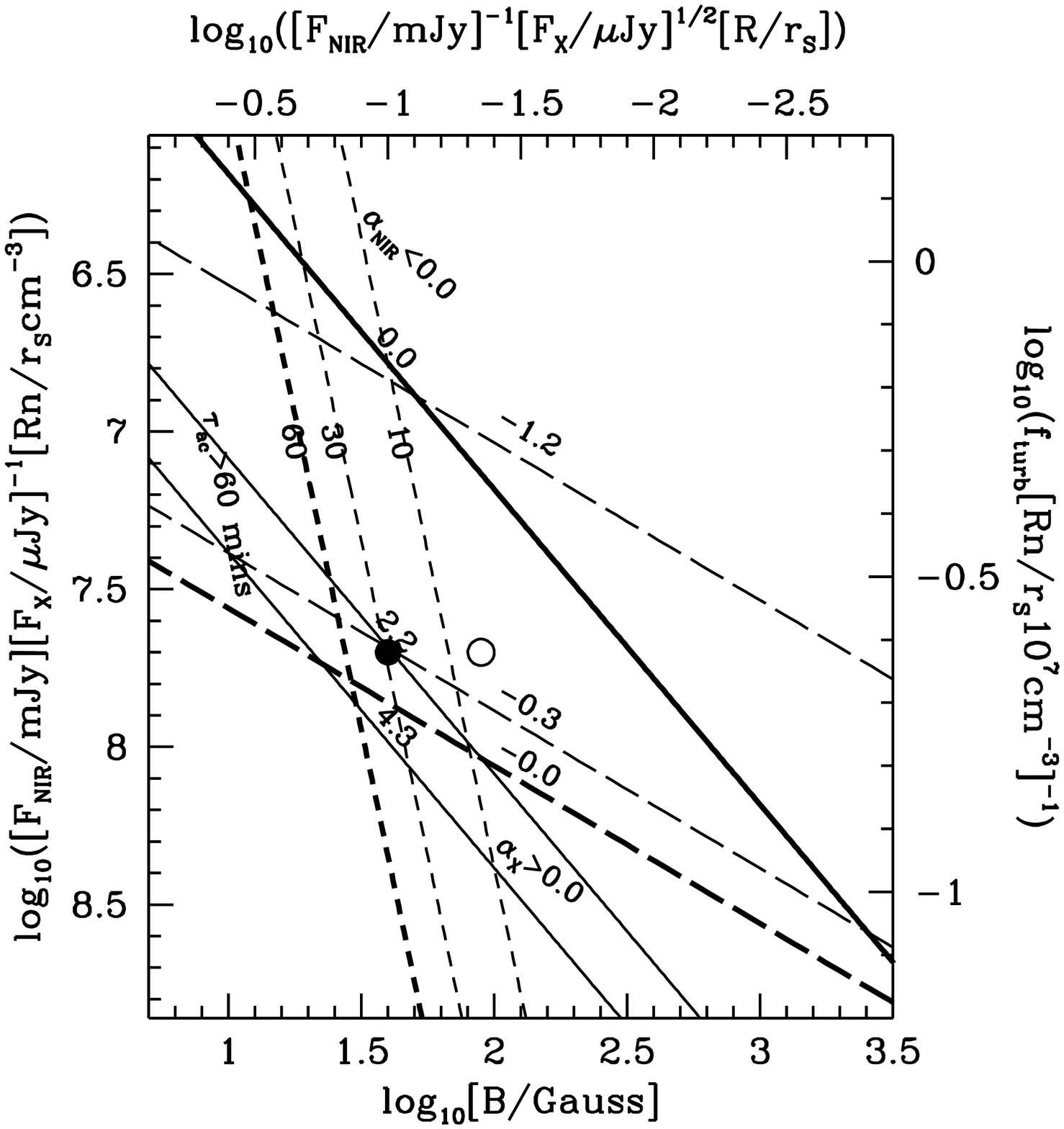}
\end{center}
\caption{
{\it Left:} Constraints on $B$ and $\gamma_{cr}$ of an SSC model for the NIR and 
X-ray flares in Sagittarius A*. The lower-left portion of the diagram is excluded 
because the SSC photons have a very soft spectrum there. The upper-right portion 
corresponds to models with very hard NIR spectra. For the region with small $B$, 
the acceleration time can be much longer than the rise time of the flare. The 
three steep dashed lines give the acceleration time (indicated), which is four times shorter than 
the synchrotron time at $\gamma_{cr}$. Viable models are therefore located in the central 
region bounded by the thick lines. The solid and long-dashed lines are for three different 
synchrotron and SSC spectra 
produced by electrons with an isotropic relativistic Maxwellian distribution. The corresponding 
spectral indexes $\alpha$ at $1.4\times 10^{14}$Hz and $10^{18}$Hz ($\sim 4.1$ keV) are indicated. 
For electron distributions different from the Maxwellian one, these lines will shift accordingly. 
The filled and the open circle are for the models indicated by the thick solid and dashed line in 
right panel of Figure \ref{fig2.ps}, respectively. {\it right:} Same as the left panel, but for 
constraints on $B$, $R$, $n$, and $f_{\rm turb}$ from the observed NIR ($\alpha_{\rm NIR}$) and 
X-ray ($\alpha_{\rm X}$) spectral indexes, the flux density at $1.4\times10^{14}$ Hz F$_{\rm NIR}$ 
and $10^{18}$Hz F$_{\rm X}$. The source is approximated as a uniform sphere. 
} 
\label{fig1.ps}
\end{figure}

Before describing the acceleration model we discuss how the existing observations limit the possible 
ranges of these parameters. As mentioned above the very steep NIR spectrum requires the cutoff 
frequency of the synchrotron emission $\nu_{cr} = 3eB_\bot\gamma_{cr}^2/4\pi m_e c=1.7 
\times 10^{12} (\gamma_{cr}/100)^2(B_\bot/40\;{\rm G})$ Hz in the sub-mm to NIR range (Eisenhauer et 
al. 2005). This sets an upper limit on $\gamma_{cr}^2B_\bot$. The spectrum of the SSC photons is 
also very sensitive to $B_\bot$ and $\gamma_{cr}$.  In the SA model described below, the scattering 
rate is much higher than the acceleration and energy loss rates. The electron distribution is 
isotropic. We therefore consider the pitch angle averaged results. The solid and long-dashed lines 
in Figure \ref{fig1.ps} ({\it left}) represent three spectra produced by electrons with a Maxwellian 
distribution with the spectral indexes $\alpha \equiv-{\rm d}\ln{(\nu F_\nu)}/{\rm d}\ln({\nu})$ at 
$\nu = 1.4\times 10^{14}$ Hz ($\alpha_{\rm NIR}$) and $\nu = 10^{18}$ Hz ($\alpha_{\rm X}$) 
indicated by the label on the lines. The change in $\alpha$ as one moves about the $B$-$\gamma_{cr}$ 
plane is due solely to the dependence of $\nu_{cr}$ on $B$ and $\gamma_{cr}$. For flares with a soft 
NIR spectrum ($\alpha_{\rm NIR}<0$) and a hard X-ray spectrum ($\alpha_{\rm X}>0$), one can exclude 
the upper-right and lower-left portions of the $B$-$\gamma_{cr}$ plane.

The acceleration time $\tau_{\rm ac}$, which is independent of energy and equal to one quarter of 
the energy loss time at $\gamma_{cr}$ in the SA model discussed below, must be shorter than or 
comparable to the flare rise time $\tau_{\rm ris}<60$ minutes, except that the source is strongly 
Doppler-boosted toward the observers. Because the luminosity of X-ray flares is usually smaller than 
that of the sub-mm to NIR flares, we can set the energy loss time equal to the synchrotron time 
$\tau_{\rm syn}(\gamma) = {9m_e^3c^5/4 e^4B^2\gamma}=\tau_0/\gamma$,
%= 18 \left[{B\over 40\,{\rm G}}\right]^{-2}\left({\gamma\over 300}\right)^{-1} 
%{\rm mins}\,,
%\label{tsyn}
%\end{equation}
where we have assumed that the source is optically thin. (The SSC cooling due to a radiation field 
with an energy density $U_{\rm syn}$ can be readily incorporated by replacing $B^2$ by an 
effective field $B_{\rm eff}^2 = B^2+8\pi U_{\rm syn}$ whenever there exists an observational 
justification.) 
%\begin{equation}
%B > 27
%\left({\gamma_{cr}\over 300}\right)^{-1/2} 
%{\rm G}\,.
%\label{timemax}
%\end{equation}
The dashed lines in Figure \ref{fig1.ps} ({\it left}) give three acceleration 
times. The left-hand side of the $B$-$\gamma_{cr}$ plane can be excluded 
because $\tau_{\rm ac}> 60$ mins there. So to produce a flare with $\alpha_{\rm NIR}>0.0$, 
$\alpha_{\rm X}<0.0$, and $\tau_{\rm ris}<60$ mins via synchrotron and SSC, the 
maximum Lorentz factor and magnetic field must be located within the central region of Figure 
\ref{fig1.ps}. In principle, simultaneous NIR and X-ray spectroscopy during the flares 
can thus directly fix $B$ and $\gamma_{cr}$.

Let us now consider how simultaneous NIR and X-ray observations with good spectral information for 
both can in addition provide us with a measurement of $R$ and the density $n$ of the 
radiating electrons. The synchrotron luminosity due to an isotropic relativistic Maxwellian electron 
population can be estimated as follows (Pacholczyk 1970)
\begin{eqnarray}
{\cal L}_{\rm syn}&=& 
{16 e^4\over 3m_e^2 c^3} {\cal N} B^2\gamma_{cr}^2 \nonumber \\ 
&=& 2.0\times 10^{36} 
\left({{\cal N}\over 10^{43}}\right) 
%\left({n\over 10^{6}\, {\rm cm}^{-3}}\right)
\left({B\over 40\, {\rm G}}\right)^2
\left({\gamma_{cr}\over 100}\right)^2 {\rm ergs\ s}^{-1}\,,
\label{lum}
\end{eqnarray}
where ${\cal N}=2N_0\gamma_{cr}^3$ is the total number of the accelerated electrons. For an 
isotropic radiation field, the corresponding X-ray emission produced via SSC is then given by
\begin{eqnarray}
{\cal L}_{\rm SSC} &=& {U_{\rm syn}\over U_B} {\cal L}_{\rm syn}\simeq {8\pi {\cal 
L}^2_{\rm syn}\over c A B^2} \nonumber  \\
&=&5.2\times 10^{35}
\left({{\cal L}_{\rm syn}\over 10^{36}\,{\rm ergs\ s}^{-1}}\right)^2
\left({B\over 40\,{\rm G}}\right)^{-2}
\left({A\over r_S^2}\right)^{-1} {\rm ergs\ s}^{-1}\,,
\label{ssc}
\end{eqnarray}
where $U_B=B^2/8\pi$ and $A$ is the magnetic field energy density and the surface 
area of the source, respectively, and ${\cal L}_{\rm syn} \simeq U_{\rm syn} c A$. From these we 
can obtain $A$ and ${\cal N}$. For a uniform spherical source, $A = 4\pi 
R^2$ and ${\cal N}=4\pi R^3n /3$, we have 
\begin{eqnarray}
R &\simeq& 0.64
\left({{\cal L}_{\rm syn}\over 10^{36}\,{\rm ergs\ s}^{-1}}\right)
\left({{\cal L}_{\rm SSC}\over 10^{35}\,{\rm ergs\ s}^{-1}}\right)^{-1/2}
\left({B\over 40\,{\rm G}}\right)^{-1} r_S\,,\label{conr} \\
n &\simeq& 4.6\times 10^6 
\left({{\cal L}_{\rm syn}\over 10^{36}\,{\rm ergs\ s}^{-1}}\right)^{-2}
\left({{\cal L}_{\rm SSC}\over 10^{35}\,{\rm ergs\ s}^{-1}}\right)^{3/2}
\left({B\over 40\,{\rm G}}\right)
\left({\gamma_{cr}\over 100}\right)^{-2} {\rm cm}^{-3}\,.
\label{conn}
\end{eqnarray}
This source size and density are typical values expected in the accretion torus of Sagittarius A* 
(LPM04), and, in general, depend on the source structure and geometry. 
The right panel of Figure \ref{fig1.ps} shows how these quantities may be read directly from 
simultaneous NIR and X-ray spectroscopic observations: From the NIR (solid) and X-ray (long-dashed) 
spectral indexes, one can locate the flare in this parameter space; The bottom axis then gives the 
magnetic field; With the NIR and X-ray flux density measurements, $R$ can be read from the top axis; 
The left and right axes then give the density $n$ of the accelerated electrons and the turbulence to 
magnetic field energy density ratio $f_{\rm turb}$, which we shall discuss in the next section. 

Flares are likely triggered by some plasma instability, or by changes in the dynamics of the 
accretion flow---for example, the dissipation of angular momentum is different above and below the 
MSO, which may lead to a strong gravitational dissipation and acceleration of electrons.
It is important to note that the total number of energetic electrons 
required to produce these bright flares can set a limit on the mass accretion rate $\dot{M}$ and 
accretion time $\tau_{\rm accr}:$ $\tau_{\rm accr} \dot{M} \ge {\cal N}m_p$, where $m_p$ is the 
proton mass. If $\tau_{\rm accr}\simeq\tau_{\rm ris}$, then 
\begin{equation}
\dot{M} \gtrsim 0.9\times 10^{16}
\left({{\cal N}\over 10^{43}}\right)
\left({\tau_{\rm ris} \over 30 {\rm mins}}\right)^{-1}
{\rm g\ s}^{-1}\,,
\label{conntot}
\end{equation}
which is consistent with the value inferred from Sagittarius A*'s quiescent state spectrum and can 
account for the observed flares for a radiation efficiency of $\sim 10\%$.

\section{Toy Model of Stochastic Electron Acceleration}
\label{model}

Let us now turn our attention to the SA of electrons and examine how this process meets the 
challenges posed by the above constraints.  Electrons can be accelerated by turbulent plasma waves 
via the transit-time damping process and cyclotron resonances. The former dominates in plasmas where 
the gas pressure is higher than the magnetic field pressure (Yan \& Lazarian 2004), and the pitch 
angle scattering rate is higher than the acceleration rate when the Alfv\`{e}n 
velocity $v_{\rm A}=B(4\pi n m_p)^{-1/2}<c$ (Schlickeiser \& Miller 1998), resulting in an 
isotropic electron distribution. The corresponding acceleration rate is proportional to 
the electron energy, giving rise to an energy independent acceleration time:
\begin{equation}
\tau_{\rm ac} = {cR\over \pi^2 v_{\rm A}^2 f_{\rm turb}}\,,
\end{equation}
which is determined by the turbulence generation length $\sim R$, $f_{\rm turb}$, and $v_{\rm A}$
(Miller, LaRosa, \& Moore 1996 \footnote{Note $\tau_{\rm ac}\equiv \gamma m_ec^2/A$, where the 
acceleration rate $A$ (not to be confused with the source surface area defined above) is given by 
Equation 2.4a in the paper.}). The SA can therefore be addressed by solving the following kinetic 
equation for the spatial and pitch angle integrated electron distribution function $N$ 
(Petrosian and Liu 2004)
\begin{equation} 
   {\partial N \over \partial t}
=  {\partial \over \partial \gamma}\left[{\partial \gamma^2N \over \partial 
\gamma} - \left(4\gamma-{4\gamma^2\tau_{\rm ac}\over \tau_0}\right) 
N\right]-{N\over T_{\rm esc}}+ \dot{Q}\,,
\end{equation}
where the source term $\dot{Q}$ and $N/T_{\rm esc}$ give the rate of particles entering and 
escaping from the acceleration site, which we identify as the emission region. 
%The importance of each term depends on the mechanism via which the plasma is excited. 
As we shall see below, the energy density of energetic electrons is
at least one order of magnitude higher than the magnetic field energy density. We therefore
envision that the plasma is excited by certain internal instability. In such a scenario, the escape
time $T_{\rm esc}$ can be much longer than the duration of the flare. 
%If $\dot{Q}$ only exists for a short period and $T_{\rm esc}$ does not change with time and energy, 
%at later time the electron distribution approaches a relativistic Maxwellian distribution 
%(Schlickeiser 1984):
%When the system reaches equilibrium, $Q=0$, and there is no
%electron flux in energy space: $(\gamma^2/ 2\tau_{\rm ac})(\partial N / \partial \gamma) - 
%(\gamma/\tau_{\rm ac}-\gamma/\tau_{\rm syn})N=0$. 
%\begin{equation}
%N(\gamma, t)=N_0\gamma^2\exp{(-a\gamma-t/T_{\rm esc})}\,,
%\label{distr}
%\end{equation}

%where we have defined Then at $\gamma_{cr}$, $\tau_{\rm ac} = 2\tau_{\rm syn}$ and
%The NIR and X-ray spectral index $\alpha$ therefore can be as small as $-1.3$ for such a model.

\begin{figure}[bht]
\begin{center}
\includegraphics[height=8.4cm, width=8.0cm]{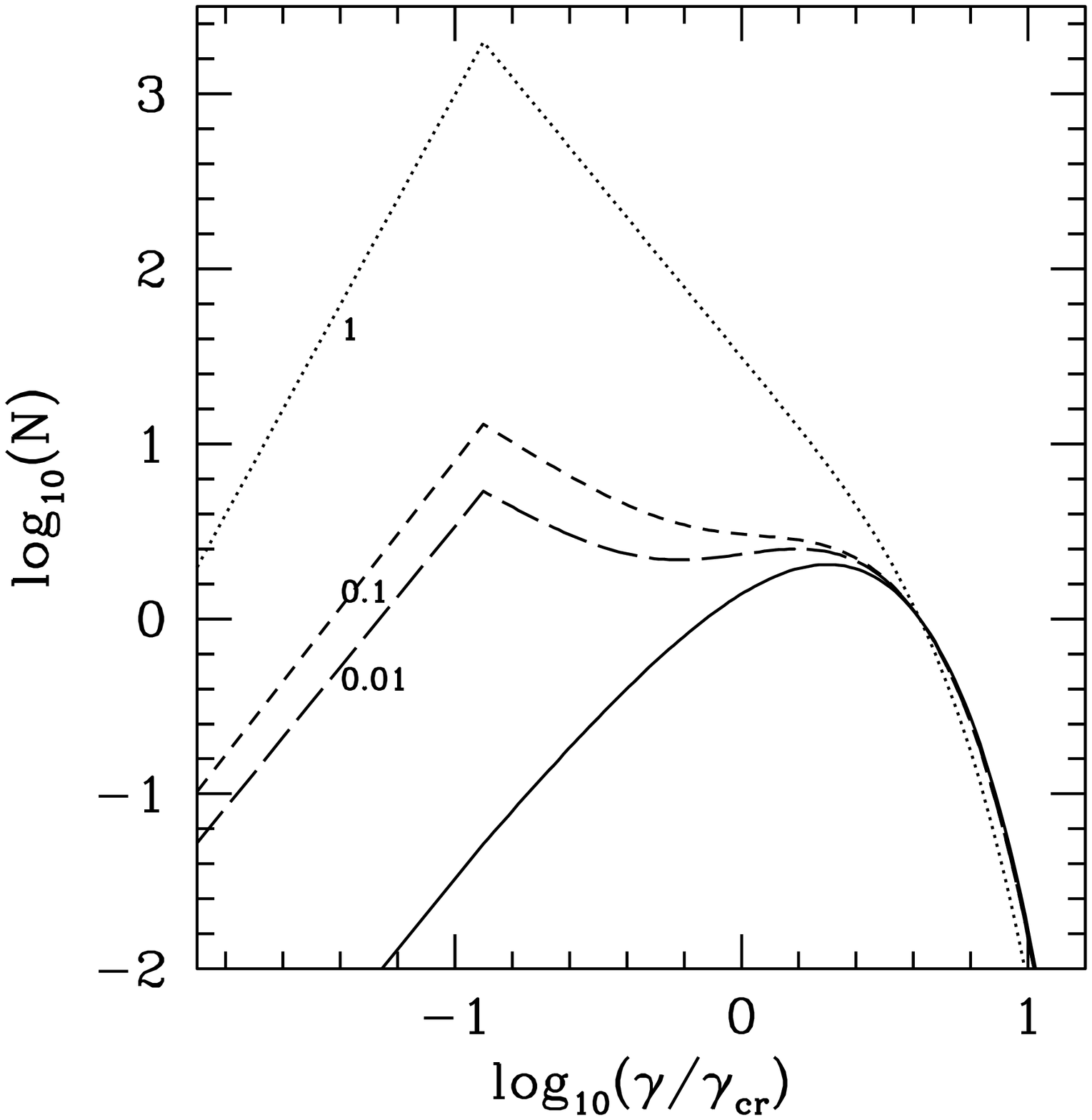}
\hspace{-0.6cm}
\includegraphics[height=8.4cm]{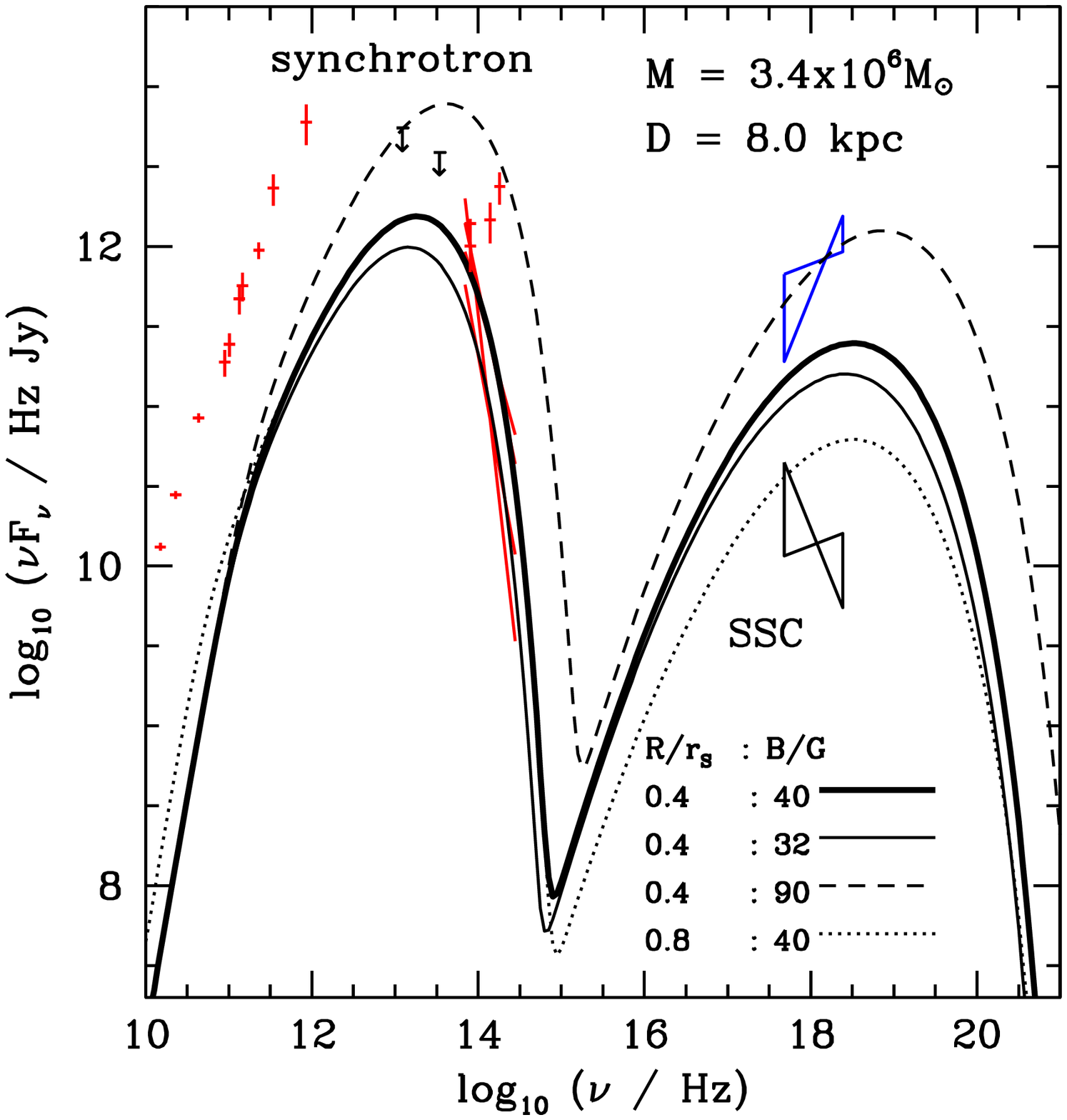}
\end{center}
\caption{{\it left:} The steady-state electron distribution normalized at $4\gamma_{cr}$
for values of $\delta$ indicated. The particles are injected at $\gamma_{\rm in}=0.13\gamma_{cr}$. 
The escape time is longer for smaller values of $\delta$, and the spectrum converges to a curve very 
close to the one for $\delta=0.01$ as $\delta\to 0$. The solid line is for the Maxwellian 
distribution. {\it right:} A fit to the NIR and X-ray flare spectrum. The parameter values include a 
Lorentz factor $\gamma_{cr}=75$ and a total number of electrons ${\cal N} = 3.8\times 10^{42}$. The 
emitting plasma is assumed to be a uniform sphere, with $R$ and $B$ as indicated. The data include 
the {\it Chandra} steady state flux (lower butterfly), and the peak X-ray spectral component of the 
27 October 2000 flare (Baganoff et al.  2001). In the NIR, the data points represent peak fluxes 
during individual events, and the crosses (barely visible below the theoretical curves) are the NIR 
spectra measured during the flare of July 15th, 2004 (Eisenhauer et al. 2005). There are 4 
theoretical curves: (thick solid) overall fit to the peak NIR emission of the July 15th event; (thin 
solid) the rising and decay phase spectrum of this NIR event; (dashed) a fit to the peak emission of 
the 27 October 2000 {\it Chandra} flare; and (dotted) is the same as the thick solid curve, except 
that $R=0.8\;r_S$.  The data points to the left of the panel correspond to the peak flux of variable 
radio and mm emission (fluctuating on a time scale of days to weeks), that originates at larger 
radii (LPM04). 
}
\label{fig2.ps}
\end{figure}

For a constant injection rate $\dot{Q}(\gamma) = \dot{Q}_0\delta(\gamma-\gamma_{\rm in})$ with 
$\gamma_{\rm in}<\gamma_{cr}$, where
\begin{equation}
\gamma_{cr} ={\tau_0\over 4\tau_{\rm ac}} = {9\pi^2m_e^3c^4v_{\rm A}^2f_{\rm turb}\over16e^4RB^2}=
30 \left({R\over r_S}\right)^{-1}
\left({n\over 10^7\,{\rm cm}^{-1}}\right)^{-1}
\left({f_{\rm turb}\over 0.1}\right)
\,,
\label{gcr}
\end{equation}
the steady-state spectrum cuts off exponentially above $\gamma_{cr}$ but 
is different from the Maxwellian one (Park \& Petrosian 1995; Schlickeiser 1984):
\begin{equation}
N(\gamma)=N_0 \gamma^{\delta+2}\exp{(-\gamma/\gamma_{cr})}\int_0^\infty 
x^{\delta-1}(1+x)^{3+\delta} \exp{(-\gamma x/\gamma_{cr})} {\rm d}x
\ \ \ \ {\rm with}\ \ \ \ \delta=\left({9\over4}+{2\tau_{\rm ac}\over T_{\rm 
esc}}\right)^{1/2}-1.5\,.
\end{equation}
%This spectrum approaches to $N_0\gamma^{-1}\exp{(-\gamma/\gamma_{cr})}$ as $\delta\to 0$.
The left panel of Figure \ref{fig2.ps} shows this distribution for $\gamma_{\rm 
in} = 0.13\gamma_{cr}$ and for several values of $\delta$ as indicated. The solid line gives the 
Maxwellian distribution. For large $T_{\rm esc}$ or small $\delta$, most of the observed 
emissions are produced by electrons near or above $\gamma_{cr}$, consequently the slope of the 
spectrum below $\gamma_{cr}$ is unimportant, and the Maxwellian distribution provides a good 
approximation of the accelerated particle spectrum. For $N(\gamma)=N_0
\gamma^{-p}\exp(-\gamma/\gamma_{cr})$, our calculations show that nearly identical 
photon spectra can be produced for different values of $p$ by adjusting $N_0$. \footnote{Note that should the 
escape time be short, i.e. when $\delta\simeq (2\tau_{\rm ac}/T_{\rm esc})^{1/2}$ is large, there could be a huge 
low energy electron population for $\gamma_{\rm in}\ll\gamma_{cr}$. Equation (\ref{gcr}) shows that 
$f_{\rm turb}\simeq 1$ is required to accelerate electrons to GeV energies.}

There are therefore four model parameters: $B$, $R$, $n$, and $f_{\rm turb}$ (or 
$\gamma_{cr}$). In the previous section, we have shown that the observed emission characteristics of 
NIR and X-ray flares set strict limits on these quantities. A detailed spectral fitting for 
individual flares can be used to measure them. The thick solid line in the right panel of Figure 
\ref{fig2.ps} fits the peak spectrum of the NIR flare of July 15, 2004, using $R = 0.4\, r_S$, $B=40$ 
Gauss, $n=1.4\times 10^7$ cm$^{-3}$, and $f_{\rm turb}=0.14$ (or $\gamma_{cr}=75$). Here the NIR 
spectrum alone determines $B\gamma_{cr}^2$ and ${\cal N}$, and the X-ray spectrum, 
which we assume to be hard and a factor of $\sim 20$ above the flux density of the quiescent state, 
constrains $R$ and $B\gamma_{cr}^4$. This model is designated by the black dot in Figure 
\ref{fig1.ps}. If the NIR flares always have very soft spectra, say with $\alpha>1.4$, then the 
magnetic field associated with these flares will fall into the narrow range $16(\tau_{\rm syn}/60{\rm 
\ mins})^{-2/3}\,{\rm G}<B<120\,{\rm G}$, and $\gamma_{cr}$ will be between 53 and $140(\tau_{\rm 
ac}/60 {\rm \ mins})^{1/3}$.

The source is self-absorbed in the sub-mm and longer wavelength range. For the source size inferred 
from the NIR flares the expected sub-mm flux will be $\sim 0.3$ Jy, which would be difficult to 
detect decisively. However, SA accompanying these flares in a large volume can produce energetic 
electrons, which produce strong mm and longer wavelengths emission {\it in situ} or in the 
process of escaping toward larger radii (Zhao et al. 2004). It is also interesting to note that the 
acceleration time $\tau_{\rm ac}$ is $27$ mins for this model, and the energy flux associated with 
the cascading Kolmogorov turbulence is $\sim f_{\rm turb}^{3/2} 4\pi R^2v_{\rm A} U_B\sim 10^{35}$ 
ergs s$^{-1}$ (Miller et al. 1996). These values are fully consistent with the flare observations.

In the right panel of Figure \ref{fig2.ps}, the thin solid line corresponds to a model with $B=38$ 
Gauss, which fits the rising and 
decay phase spectrum of the NIR flare of July 15, 2004. The dashed line fit corresponds to another 
model with $B=90$, which fits the peak flux of the X-ray flare of 27 October, 2000. All other model 
parameters have the same values as those described above. These results suggest that the variations 
in flare characteristics may be attributed to changes in the magnetic field. However, we emphasize 
that the optically thin NIR emission only depends on the total number of energetic electrons, while 
the X-ray emission also depends on the source area. To demonstrate this effect, the dotted line 
shows a spectrum produced with the area of the emission region increased by a factor of 4
(i.e., $R=0.8 r_S$. Note ${\cal N}$ is fixed). As expected, the NIR spectrum is unaffected, while 
the X-ray flux decreases by a factor of 4, as expected from equations (\ref{lum}) and (\ref{ssc}).

\section{Conclusions and Discussion}
\label{dis}

The simultaneous detection of flares in the NIR and X-ray bands suggests that there is an 
intimate connection between these emission mechanisms. The fact that NIR flares have very steep 
spectra, whereas X-ray flares always have a much flatter spectrum, rules out the 
possibility of producing the NIR and X-ray photons together via synchrotron emission. 
The simplest model for production of X-ray is SSC model, but it faces several challenges. In this 
paper we have shown how the simultaneous observation of NIR and X-ray flares may be used to determine 
the source size, the magnetic field, and the distribution of electrons, should the NIR and 
X-ray emission be produced, respectively, via synchrotron and SSC. We have shown that the 
emission characteristics suggest an electron distribution cutting off at $\gamma_{cr}\sim 100$. The 
size of the source must be a fraction of a Schwarzschild radius in order to produce 
prominent X-ray flares. The rapid rise in the NIR emission suggests a magnetic field of a few tens 
of Gauss. A lower magnetic field is required should the source be Doppler-boosted, producing flares 
with soft NIR and hard X-ray spectra. 

We have demonstrated that the SA model we proposed previously is fully 
consistent with the current observation of flares from Sagittarius A*, and with the transit-time 
damping acceleration taking the place of the parallel propagating waves we used before, the 
acceleration only depends on the turbulence energy density. However, depending on details of the 
flare excitation mechanism, the electron distribution below $\gamma_{cr}$ can be quite different.
This model therefore can be a powerful tool in probing the plasma energization and particle 
acceleration processes near the event horizon of the black hole. 

The rest frame acceleration time is slightly longer than the typically observed rise time. This 
suggests that a time-dependent treatment of the SA model is necessary unless the emitting plasma is 
Doppler-boosted by a boosting factor $\Gamma\gg \tau_{\rm ac}/\tau_{\rm ris}$. However, our main 
conclusions still hold---current observations 
cannot yet distinguish the subtle difference. Nevertheless, such a calculation is clearly warranted 
with the acquisition of new data, particularly as flares continue to be observed simultaneously 
across the spectrum. 

It is also interesting to note that should the magnetic field be anchored to a slow large scale flow, 
it will exert a stress on the plasma. This stress causes transport of the angular momentum of the 
emitting plasma. In a Newtonian potential where the angular momentum density in a Keplerian orbit 
is given by $L = m_pn (GMr)^{1/2}$, we have
$m_pn (GM)^{1/2} r^{-1/2} v_r \sim (\nabla\times {\bf B})\cdot{\bf B} r \simeq B^2r/R \,,$
or
$$v_r \sim 2.3\times 10^{10} 
\left({R\over r_S}\right)^{-1}
\left({r\over 3\,r_S}\right)^{3/2}
\left({B\over 40\,{\rm G}}\right)^2
\left({n\over 10^7\,{\rm cm}^{-3}}\right)^{-1}
{\rm cm\ s}^{-1}\,,$$
which is comparable to the Keplerian velocity at $r= 3r_S$ and is much higher than the radial 
velocity $v_r \sim 10^8$ cm s$^{-1}$ suggested by the chirping behavior of the X-ray and NIR flares, 
indicating that the variation length scale of this external magnetic field should be much longer 
than $R$, or that most of the magnetic field is 
generated within the plasma itself. Should neither of these scenarios be viable, a large black hole 
spin would then be required.

This model is also in line with our previous study of Sagittarius A*'s emission in the 
quiescent state (LPM04), and with our study of proton and electron 
acceleration on larger scales ($\sim 20 r_S$; Liu, Melia, and Petrosian 2005). Clearly 
an appropriate modeling of the magnetic field structure in the accretion flow of 
Sagittarius A*, and in any possible outflow, is required in order for us to have a 
complete understanding of this intriguing object. With it, we should hope to test its 
predicted correlated variability over a broad range in frequencies.

\acknowledgments

We thank Fred Baganoff for providing the most recent results of their multi-wavelength 
monitoring of Sagittarius A* and Huirong Yan for useful discussion. This research was partially 
supported by NSF grant ATM-0312344, NASA grants NAG5-12111, NAG5 11918-1 (at 
Stanford), and NSF grant AST-0402502 (at Arizona). FM is very grateful to the University of Melbourne 
for its support (through a Miegunyah Fellowship).

\end{document}